\documentclass[5p,twocolumn]{elsarticle}
\usepackage{graphicx}
\usepackage{float}
\usepackage{subfig}
\usepackage{lineno}
\usepackage{amsmath}
\usepackage{amssymb}
\usepackage{fullpage}
\journal{Nuclear Instruments and Methods A}

\newcommand{\Utte}{$^{238}$U}
\newcommand{\Th}{$^{232}$Th}
\newcommand{\Ctft}{$^{252}$Cf}
\newcommand{\Het}{$^3$He}
\newcommand{\Hef}{$^4$He}

\begin{document}
\begin{frontmatter}
\title{Fast Neutron Detection with a Segmented Spectrometer}

\author[a,b]{T.~J.~Langford\corref{cor1}\fnref{tLangford}}
\fntext[tLangford]{Present Address: Yale University, New Haven, CT 06511 USA}
\address[a]{Department of Physics, University of Maryland, College Park, MD 20742 USA}
\address[b]{Institute for Research in Electronics and Applied Physics, University of Maryland, College Park, MD 20742 USA}
\author[c]{C.~D.~Bass\fnref{cBass}}
\address[c]{National Institute of Standards and Technology, Gaithersburg, MD 20899 USA}
\fntext[cBass]{Present Address: Le Moyne College, 1419 Salt Springs Rd, Syracuse, NY 13214 USA}
\author[a]{E.~J.~Beise}
\author[a]{H.~Breuer}
\author[a]{D.~K.~Erwin}
\author[c]{C.~R.~Heimbach}
\author[c]{J.~S.~Nico}
\cortext[cor1]{Corresponding Author: thomas.langford@yale.edu} 

\date{\today}

\begin{abstract}

A fast neutron spectrometer consisting of segmented plastic scintillator and \Het\ proportional counters was constructed for the measurement of neutrons in the energy range 1\,MeV to 200\,MeV. We discuss its design, principles of operation, and the method of analysis. The detector is capable of observing very low neutron fluxes in the presence of ambient gamma background and does not require scintillator pulse-shape discrimination. The spectrometer was characterized for its energy response in fast neutron fields of 2.5\,MeV, 14\,MeV, and the results are compared with Monte Carlo simulations. Measurements of the fast neutron flux and energy response at 120\,m above sea-level (39.130$^{\circ}$\,N, 77.218$^{\circ}$\,W) and at a depth of 560\,m in a limestone mine are presented.  Finally, the design of a spectrometer with improved sensitivity and energy resolution is discussed.

\end{abstract}

\end{frontmatter}


\section{Introduction}
\label{sec:Intro}

To minimize the influence of ambient backgrounds resulting from cosmic rays and natural radioactivity, many physics researchers site their experiments in deep underground laboratories~\cite{Formaggio04}. Regardless, there will still exist fast neutrons that can strike nuclei and produce recoils that may mimic signatures of rare events. Alternatively, the neutrons may activate radioactive isotopes in detector material producing background in these low event rate experiments~\cite {Gaitskell04,Angloher2012,Aprile2012,Agnese2014,Akerib2014,Abdurashitov2009, McKinsey05, Aharmim07,Elliott02,Arnaboldi2004, Aalseth2011,Ackerman2011,Agostini2013}.  An improved knowledge of the fast neutron energy spectrum and flux in the varied environments of these experiments would enable better optimization of efforts to identify and minimize such backgrounds.

Fast neutrons encountered in underground laboratories arise primarily through two mechanisms, muon-induced spallation and naturally-occurring radioactivity in the local environment. Neutrons from radioactivity are predominantly generated by spontaneous fission in the  \Utte\ and \Th\ decay chains  and ($\alpha$,$n$) reactions in surrounding material. These neutrons typically have energies of a few MeV and range up to about 15~MeV.  Neutrons from the local environment have fluxes that are related to the uranium and thorium content of the surrounding rock and are therefore independent of the laboratory depth. Neutrons of much higher energies (up to a few GeV) are produced from cosmic-ray muon-induced spallation reactions~\cite{Delorme1995,Mei2006}. These neutrons range from a few MeV to many GeV, and their flux and spectrum vary with the amount and composition of the overburden.

For many underground facilities, there is insufficient knowledge of the underground flux and spectrum at the lower energies, and the situation is typically poorer at higher energies.  The need for understanding fast neutron backgrounds has outpaced the ability to perform such measurements~\cite{Mei2006}.  Experimenters have become reliant on Monte Carlo modeling as their best approach to quantifying neutron fluxes as a function of energy~\cite{Wulandari2004, Marino2007}, particularly in the high energy regime.  In many cases, the models lack experimental benchmarks with which the results can be verified.  The problem becomes even more complicated when one considers the secondary fast neutrons that can be created from the initial spallation neutrons.  It is clear to the community of researchers in underground physics that reliable measurements of both the fluxes and energy spectra are essential to understanding their backgrounds~\cite{Guiseppe2012}. A number of collaborations address this concern by measuring fast neutron parameters directly at their experimental facility~\cite{Abdurashitov2002,Bonardi2010,Cooper2011,Bellini2011,Zhang2013138,Jordan2013,Park2013,Villano2013} and by benchmarking the codes that they use for simulations~\cite{Wulandari2004, Marino2007}. 

In this paper, we present the design and results of a Fast Neutron Spectrometer (FaNS-1) whose purpose is to enable energy and flux measurements of very low levels of fast neutrons with good energy resolution. The spectrometer consists of segmented plastic scintillator bars and \Het\ proportional counters~\cite{Langford2010,LangfordThesis}. The proton recoil in the scintillator bars permits energy reconstruction, and the capture of the delayed, thermalized neutron improves event identification without requiring pulse-shape discrimination. It has a dynamic range from approximately 0.5\,MeV to 200\,MeV, which allows the determination its energy response to both terrestrial and cosmic sources. These results can be compared with simulations to help improve the accuracy their predictions.
 
In Section \ref{sec:Detection} the principle of capture gated spectroscopy is briefly reviewed along with a method to restore energy resolution through segmenting the detector. The design of the FaNS-1 spectrometer is presented in Section~\ref{sec:fans} along with the electronics, data acquisition system, and analysis method. Results from characterization studies performed in 2.5\,MeV and and 14\,MeV neutron fields are given in Section \ref{sec:DetCharac}. In Section~\ref{sec:surface}  a measurement of the fast neutron response generated at the Earth's surface is presented. The results from measuring the fast neutron response in the Kimballton Underground Research Facility (KURF)~\cite{Finnerty2010} are given in Section~\ref{sec:kurf}. These measurements are summarized in Section~\ref{sec:summary} along with brief description of a second spectrometer under construction that will have a higher efficiency and a sensitivity over a broader energy range.

\section{Neutron Detection with Plastic Scintillator and \Het\ Proportional Counters}
\label{sec:Detection}

\subsection{Capture-gated neutron detection}

\begin{figure}
\begin{center}
\includegraphics[scale=0.7]{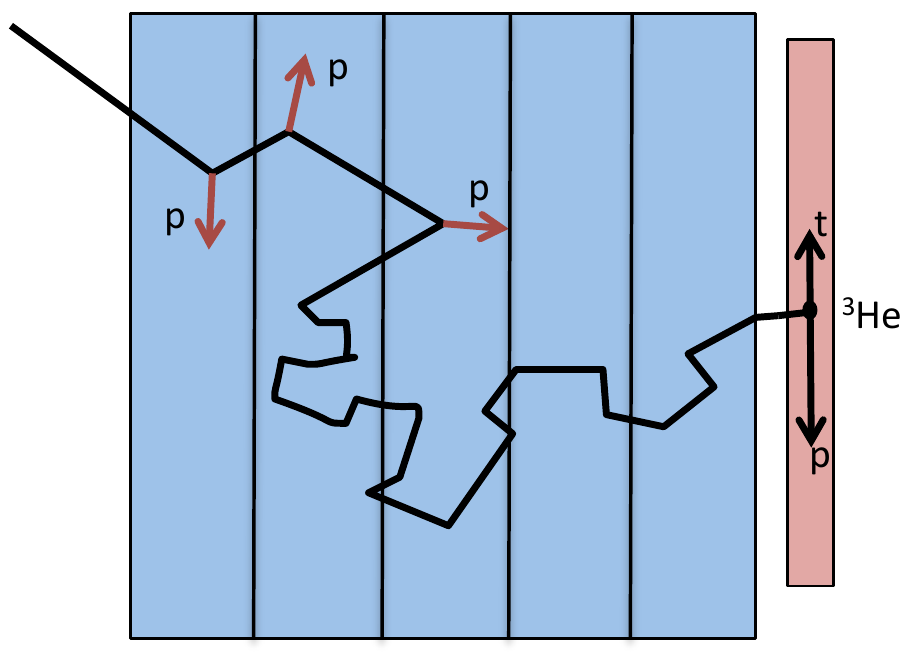}
\caption{A schematic representation of a fast neutron interacting in a block of scintillator and captured within a \Het\ proportional counter. The first few scatters create recoil protons with enough energy to be measured above a threshold. After the neutron thermalizes, it diffuses within the medium until capturing on a \Het\ nucleus. Segmentation of the detector permits reconstruction of the incident energy of the neutron, despite the nonlinear light response.}
\label{fig:multScatter}
\end{center}
\end{figure}

The detection method employed here is known as capture-gated spectroscopy~\cite{Drake1986,Czirr1989,Abdurashitov1997,Czirr2002}. In brief, the idea involves using an organic scintillator to detect fast neutrons through their recoil interaction with protons in the scintillator. The neutrons that thermalize and are captured produce a signal indicating that the recoil event was due to a neutron, as illustrated in Fig.~\ref{fig:multScatter}. This capture serves to discriminate against uncorrelated background events. It also indicates that there was full deposition of neutron's energy in the scintillator bars, thus permitting the reconstruction of the initial neutron energy.

The FaNS-1 spectrometer uses \Het\  in proportional counters as the capture agent although Gd~\cite{Pawelczak2011} is a common agent and other efforts have focused on the use of $^{10}$B~\cite{Aoyama1993,Normand2002} and $^6$Li~\cite{Fisher2011,Bass2012}.  $^3$He has two important advantages in that the reaction produces two charged particles in the final state rather than gamma rays, thus making it easy to identify, and it has a large thermal neutron cross section.  A neutron is captured by a $^3$He nucleus, resulting in a proton and a triton

\begin{equation}
^3He + n \rightarrow p + t + 764 \; \textrm{keV.}
\end{equation}

\noindent The energy is shared between the two reaction products. Purely via kinematics, the proton receives 573\,keV and the triton receives 191\,keV. 

The use of separate detectors for the proton recoil signals and the neutron capture signals has the benefit of being able to distinguish between the two~\cite{Abdurashitov1997}. Because the neutron capture always occurs after the recoil, we look for signals which have the reverse time order as a measure of the uncorrelated coincidences between the scintillator and the proportional counters. For the low rate of capture events in this work, the uncorrelated coincidences are uniformly distributed in time while the signal from a neutron that thermalizes and captures occur at short times characterized by the neutron diffusion time and the scintillator geometry. With the resulting timing spectrum, a proper background subtraction can be applied to the data. This will be discussed further in Section~\ref{sec:DetCharac}.

\subsection{Energy reconstruction and response}
\label{subsec:energy}

When a neutron undergoes thermalization in an organic scintillating medium, the main mechanism of energy loss is scattering from protons. As it does so, proton recoils yield light through the scintillation process. For incident gamma-rays, the light-output response function is linear over a wide range of energies, but that is not the case for heavy charged particles. The light output $L$ from a neutron that deposits all of its energy in a single scatter, $E^{1s}$, is not the same as one that deposits all of its energy through multiple scatterings. For example, if one considers two scatters with $E_a^{2s} + E_b^{2s} = E^{1s}$, one finds specifically that  

\begin{equation}
L\left(E_a^{2s} + E_b^{2s} \right) < L\left(E^{1s}\right).
\end{equation}

\noindent This is due to the non-proportional energy-to-light conversion where the ratio of light to energy decreases with proton energy, leading  to a deficit in the total light collected  in multiple scattering events as compared to single scattering events. For a single scintillator detector, the degree of multiple scattering is not known and thus reconstructing the kinetic energy of a stopped neutron leads to  poor energy resolution~\cite{Aoyama1993}. 

This problem can be mitigated by separating the occurrence of multiple scatters into separate segments, as shown in Fig.~\ref{fig:multScatter}. The energy transfer in each individual scatter can be reconstructed according to the light response function of the scintillator and combined to properly reconstruct the energy of the incident neutron~\cite{Abdurashitov2002a,Bowden2009}. Ideally, the segments should be small enough so that on average a neutron makes not more than one interaction per segment, but one must balance enhanced resolution against detector complexity. In a practical implementation, the segment size will be such that there is often more than one scatter within the segment, but the resolution is still significantly improved when compared with a detector of the same volume but without segmentation.

Another critical factor in obtaining the correct energy is understanding the specific light response as a function of the stopped particle energy and type of scintillator that was used. Much work has been done studying the effective light response curves for different types of liquid scintillator, including NE-213\footnote{Certain commercial equipment, instruments, or materials are identified in this paper in order to specify the experimental procedure adequately. Such identification is not intended to imply recommendation or endorsement by the National Institute of Standards and Technology, nor is it intended to imply that the materials or equipment identified are necessarily the best available for the purpose.} and BC-501~\cite{Nakao1995,Verbinski1968}. If the specific energy loss is known for the particular material, it is possible to calculate the nonlinear light response function~\cite{Birks1951, Craun1970,ORielly1996}.  The method is based upon the concept that the quenching of light stems from recombination of electrons and ions in the scintillator. For heavy charged particles, which have high specific energy loss and deposit most of their energy in a very small range, the quenching is enhanced by the increased density of electrons and ions. Some pairs recombine, and therefore do not produce light.

A fitting function of the following form was derived in Ref.~\cite{Craun1970} that produces the light response of protons
\begin{eqnarray}
\label{eqn:Birks}
\nonumber
dL/dx &=&	 S\times (dE/dx)\times[1 + kB\times(dE/dx)	\\
	&  &	\mbox{}+ C\times(dE/dx)^2]^{-1},
\end{eqnarray}

\noindent where $dL$ is the light produced in path length $dx$, $dE/dx$ is the specific energy loss of a particle with kinetic energy $E$, and $S$ is a scintillation efficiency determined through gamma calibration.  $kB$ and $C$ are adjustable parameters that were empirically determined to fit light response measurements in Ref.~\cite{Madey1978} for NE-102 polyvinyltoluene plastic scintillator (an equivalent to the BC-400 used in this work). The values obtained are $kB = 0.0085\,\rm{g/cm^{2} \,MeV^{-1}}$ and $C = 1\times 10^{-6}\rm\,{(g/cm^{2}\,MeV^{-1} )^{2}}$. The specific energy loss used came from the NIST database for the stopping power and range of protons and alphas in polyvinyltoluene~\cite{pstar}. This technique provides a smoothly varying function that covers the broad range of neutron energies used in this work. Recent work~\cite{Reichhart2012} has shown that there may be significant deviations from fits of Ref.~\cite{Birks1951} for plastic scintillator at low-energy recoils. This would effect both the simulated and reconstructed energies in the detector. By measuring and simulating the detector response to monoenergetic neutron sources, it is possible to test the validity of this determination for the light response.

For this work, the term ``deposited energy'' is used to describe the energy response generated by the stopped neutrons in the FaNS-1 detector. The term is defined as the neutron kinetic energy that is converted to ionization in the plastic scintillator. For many neutrons this will represent the full kinetic energy, but there exist several effects that cause distortions in the measured spectra. In addition to the nonlinear light output of the scintillator from protons, there are effects such as a greatly reduced light output for particles heavier than protons, multiple scattering in a single scintillator, and scattering in non-scintillating material. Of particular concern are carbon interactions, elastic and inelastic, which absorb energy without producing significant light. These reactions combine to widen and distort the measured peak, generally shifting the response to lower energies. If the these effects did not contribute to detected events, the resulting energy deposit spectra would represent the true neutron kinetic energy spectra. 

\section{FaNS-1 Detector}\label{sec:fans}
\subsection{FaNS-1 Design and Data Acquisition System}
\label{subsec:Design}

\begin{figure}
\begin{center}
\includegraphics[scale=.35]{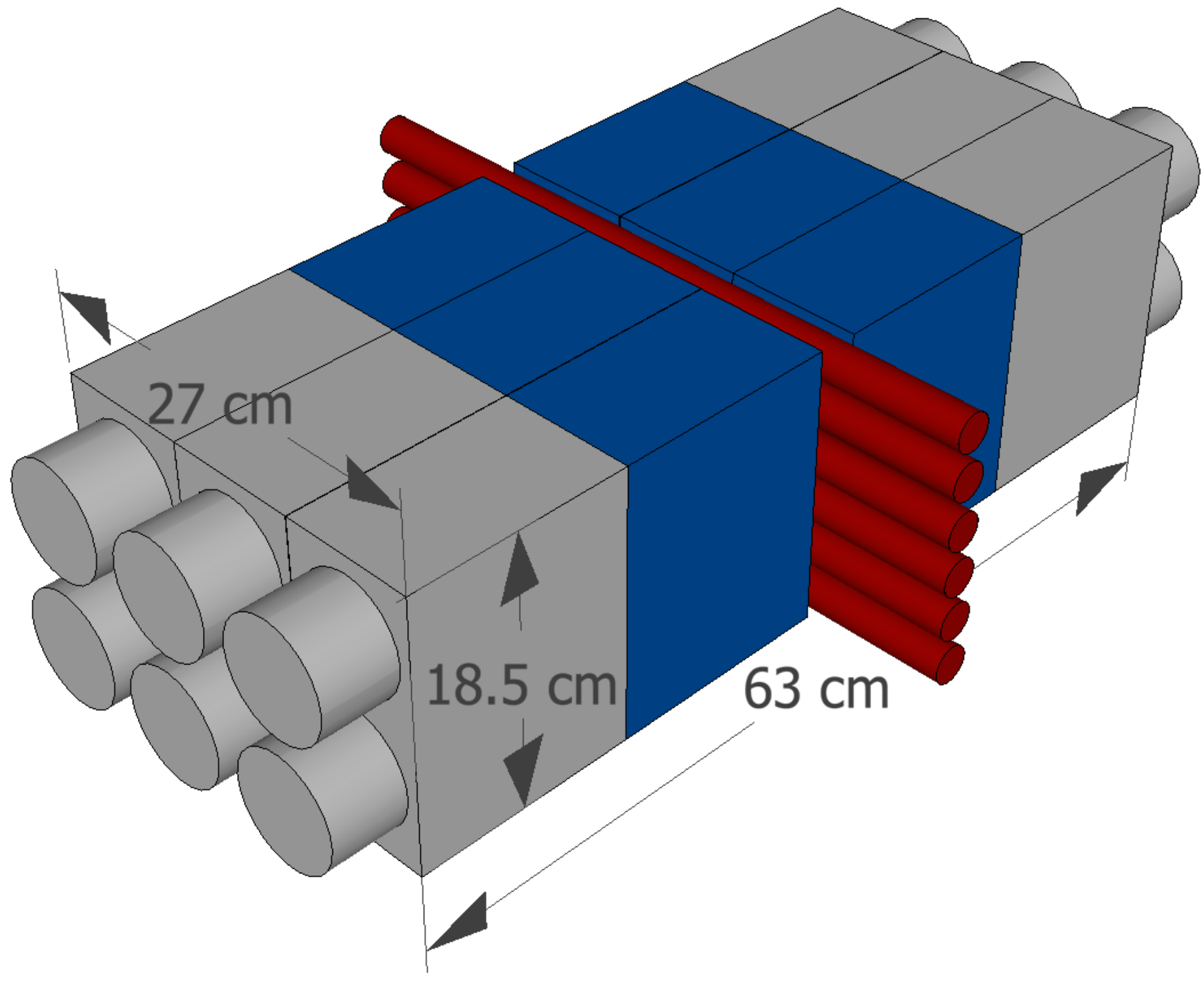}
\caption{An illustration of the FaNS-1 detector (color online). The regions shaded blue are scintillator and the grey portions are lucite light guide.  The six \Het\ proportional tubes are red and shown in the center. The PMTs are mounted to the cylindrical light guides with silicone potting.}
\label{blockPic}
\end{center}
\end{figure}

The design of the detector took advantage of segmentation and capture-gating to achieve good energy resolution and good background rejection. The detector consists of six \Het\ proportional counters and six segments of plastic scintillator, arranged as illustrated in Fig.~\ref{blockPic}. Each segment is a 9.0\,cm $\times$ 18.5\,cm $\times$ 15.0\,cm  section of BC-400 scintillator and an identical section of light guide that is glued to the scintillator. Each scintillator segment has an active volume of 2.5\,L, yielding a total active volume of 15.0\,L. Two 7.62\,cm photomultiplier tubes (PMT) are mounted at the end of each light guide via 5\,cm long cylindrical light guides. Each segment is wrapped in aluminized Mylar and made light tight with black electrical tape. This optical decoupling prevents crosstalk between the segments. The segments were repurposed from another experiment~\cite{Raue1996}, and no structural changes were made to them.

The \Het\ proportional counters were manufactured by Reuter Stokes and have a 2.54\,cm diameter aluminum body and an active length of 46\,cm. They contain a mixture of 4.0\,bar of \Het\ and 1.1\,bar of natural krypton, which increases the stopping power of the gas for heavy, charged particles. This reduces wall effects where only one of the final state products of the neutron capture deposits its full energy~\cite{Knoll2000}.  The krypton has a negligible effect on the neutron detection efficiency due to its low interaction cross section.

The detector is housed in two banks, one on either side of the row of six proportional counters. This arrangement is a convenient geometry to use given the fixed size of the scintillator segments and proportional counters and the limited flexibility in their arrangement. The housing ensures that the detectors will remain in the same location relative to each other and provides a reproducible configuration. The detector was covered with flexible, 3-mm thick, boron-loaded silicone rubber to shield from ambient thermal neutrons that would contribute to the uncorrelated background rate.

\begin{figure}
\begin{center}
\includegraphics[scale=.12]{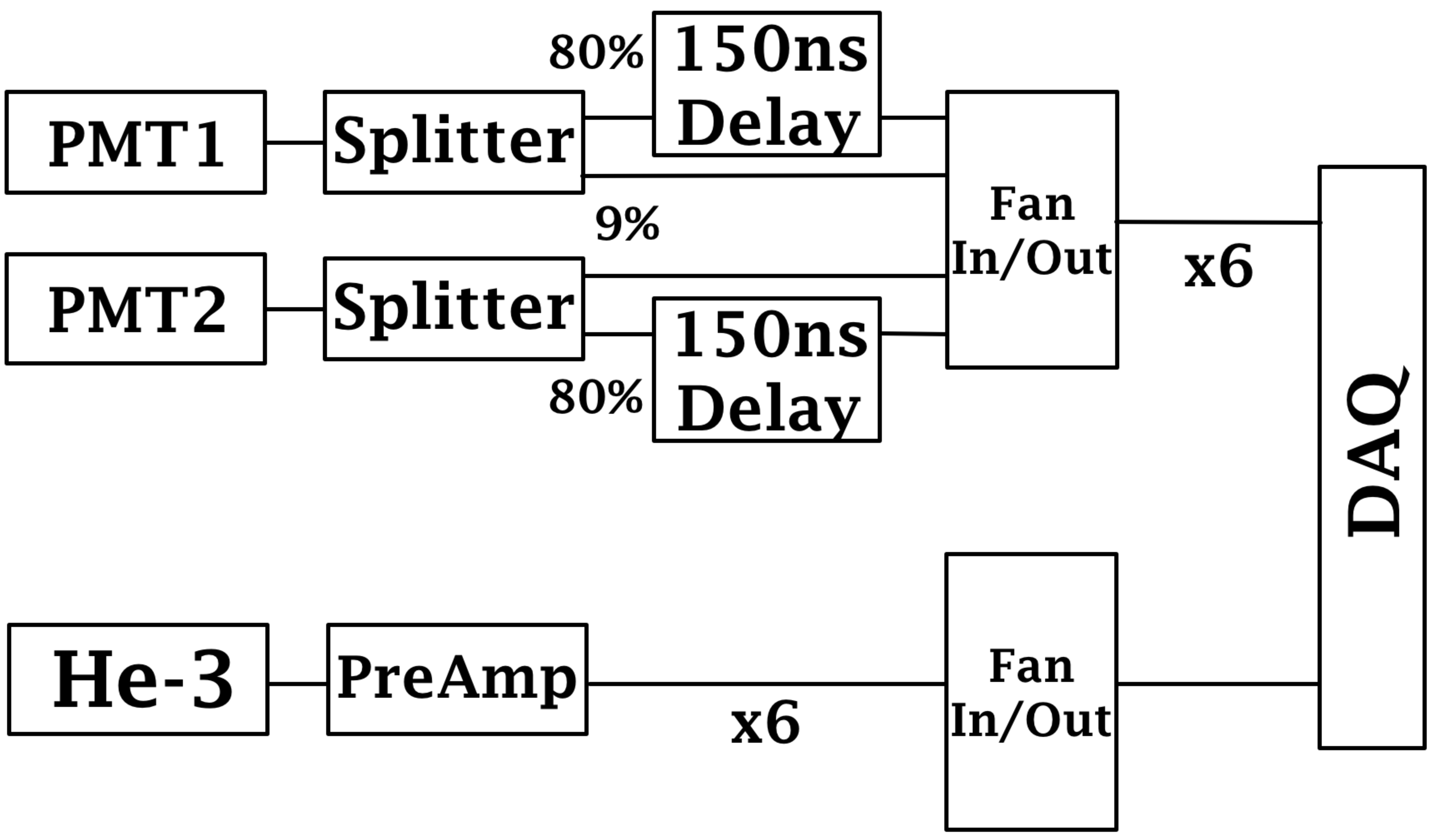}
\caption{A block diagram of the trigger electronics.  Each PMT signal from a scintillator segment is asymmetrically split; one signal is delayed and then summed with the undelayed attenuated signal and input into one channel of the waveform digitizer. Approximately 11\,\% of the initial signal is lost through the splitter box.  The data acquisition system (DAQ) is triggered internally by the \Het\ signal. }
\label{fig:SplitterCircuit}
\end{center}
\end{figure}

\begin{figure}[t]
\begin{center}
 \includegraphics[width=.45\textwidth]{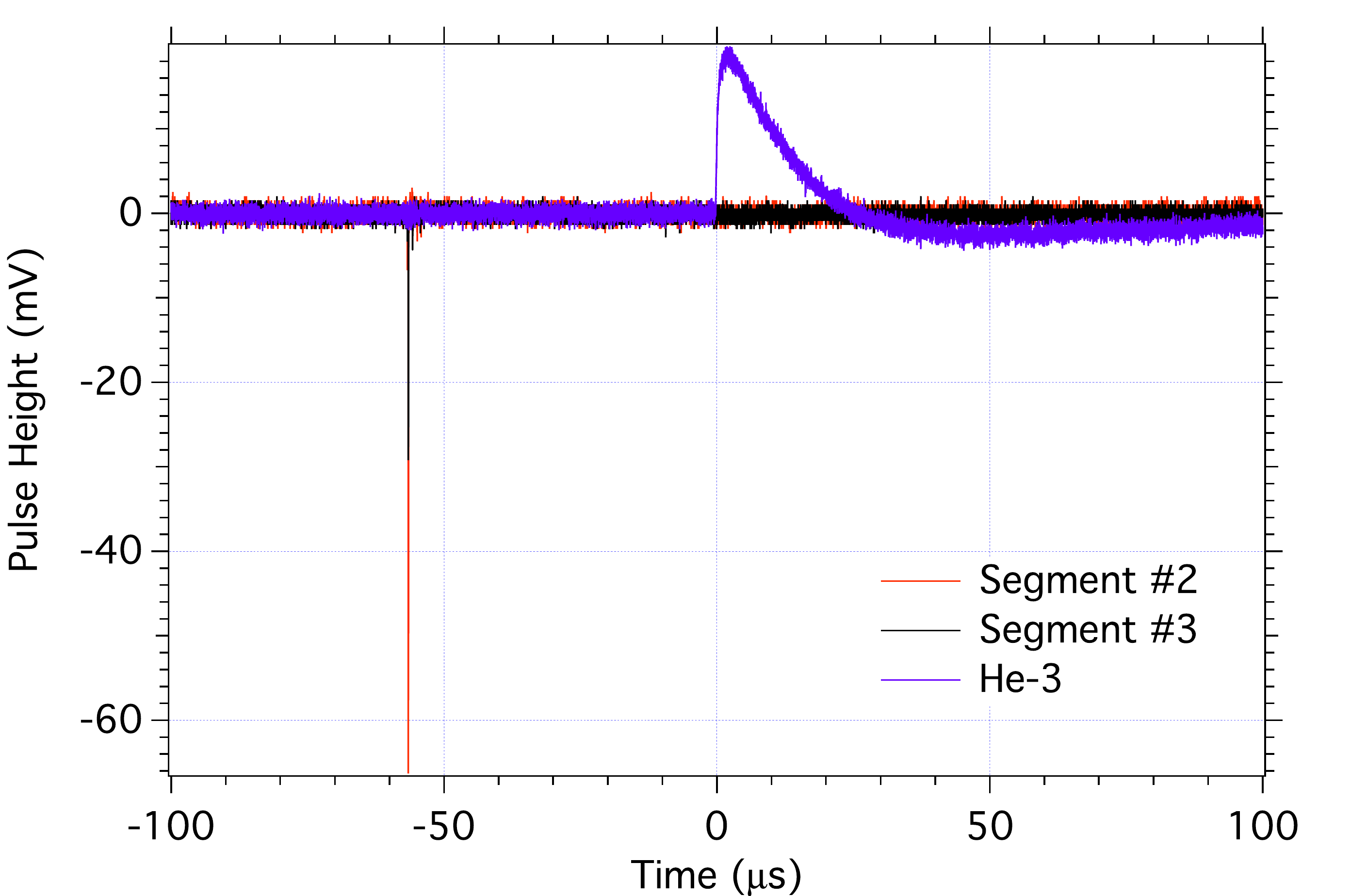}  \\
 \includegraphics[width=.45\textwidth]{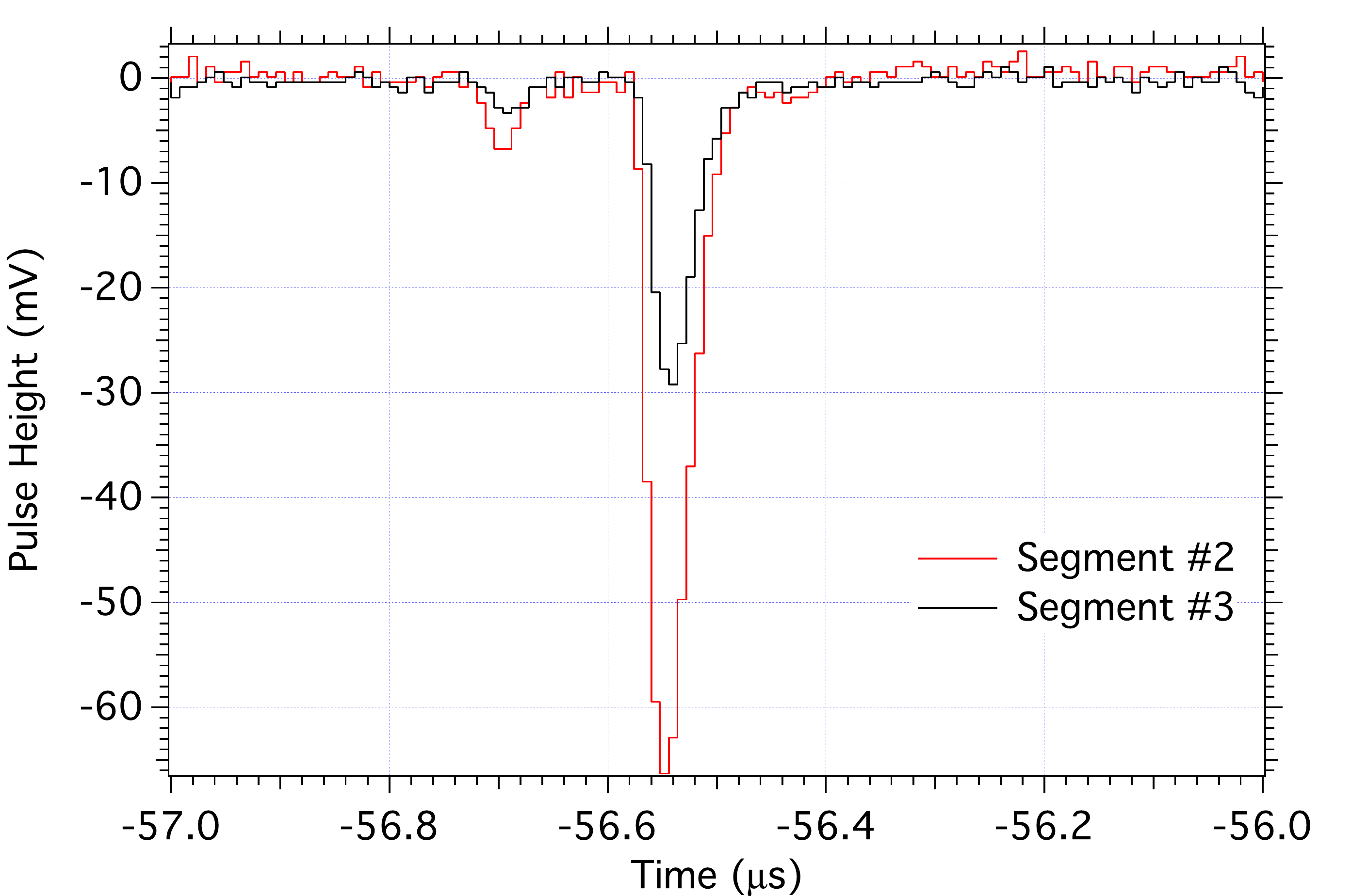}
\caption{Top: A neutron event candidate (color online). The red and black traces are digitized PMT signals from two scintillator segments, and the blue trace is the signal from a \Het\ proportional counter. Note that the scintillator signals come before the \Het\ proportional counter signal from the capture of a thermalized neutron. Bottom: An expanded view of the two separate scintillator signals. Note that they are coincident in time, indicating that they most likely arose from the same fast neutron. The delayed full and prompt attenuated signals are both apparent.}
\label{sampleTrace}
\end{center}
\end{figure}

The signal from each PMT was sent through an asymmetric splitter circuit, shown in Fig.~\ref{fig:SplitterCircuit}. The signal is passed through a passive resistor chain that produces two signals, a full-amplitude signal and an attenuated signal, which are a factor of nine different in pulse height.  This approach allows for both good resolution at low energy and a large dynamic range~\cite{Breuer13}.
The large amplitude signal was delayed by about 150\,ns, and the two signals for a given segment were summed in a linear fan-in/fan-out. The two PMT signals from each segment were summed together to reduce the number of channels needed for acquisition. The PMT and \Het\ proportional counter signals were recorded by an 8-channel, 125\,MSample/s, 12-bit PCI-based waveform digitizer. Examples of the resulting traces are shown in Fig.~\ref{sampleTrace}.

The six \Het\  proportional counters were biased through preamplifiers and did not have shaping amplification.  All six individual preamplifier signals were combined in one fan in/out module, and the output went into one channel of the waveform digitizer. The digitization of the \Het\ signals allowed the extraction of risetime and amplitude information, which in turn permitted some neutron/alpha discrimination~\cite{Langford2013}. This proved essential for achieving high sensitivity in environments where the neutron flux was very low. All of the proportional counters were tested individually and gain matched to within $5\,\%$.  

The data acquisition system triggered on a proportional counter signal, and typically $400\, \mu$s traces of all channels were digitized. Digitizing all channels allows us to set the PMT thresholds in software analysis, rather than hardware. The trigger window includes time before and after a prompt coincidence (i.e., an event where the PMT and proportional counter signals occurs nearly simultaneously); this permits  a real-time measurement of the random background of uncorrelated coincidences, which manifest themselves in events where the proportional counter trigger comes before a PMT signal. Figure~\ref{sampleTrace} shows a sample event in which a neutron interacts in two segments of the scintillator before capturing on a \Het\ nucleus. 

The energy and timing information for each event are extracted from the digitized signals in software. Both the full and attenuated signals, shown in Fig.~\ref{sampleTrace}, are analyzed. If the full signal saturates the data acquisition, the pulse height is taken from the attenuated signal.  As a safety margin, the cross-over point from full to attenuated is taken as 80\,\% of the saturation point. 

\subsection{Analysis}
\label{sec:Analysis}

The digitized waveforms were stored for offline analysis. PMT signals were integrated to determine the number of photoelectrons detected. Gamma-ray source calibrations were performed with $^{137}$Cs and $^{60}$Co and provided the conversion from photoelectrons into energy. Source availability in the detector locations limited the ability to calibrate at multiple energies. During the long running of the detector underground, the PMT gains drifted. To account for the drift, the data acquisition system was configured to trigger on any individual scintillator segment for a short period every hour, thus yielding ambient gamma spectra for each segment. These spectra included discernible $^{40}$K and $^{208}$Tl gamma lines from the surrounding rock. The Compton edge from the 2.6\,MeV $^{208}$Tl gamma was tracked over time, providing a continuous calibration reference. 

\begin{figure}[t]
\begin{center}
 \includegraphics[width=.45\textwidth]{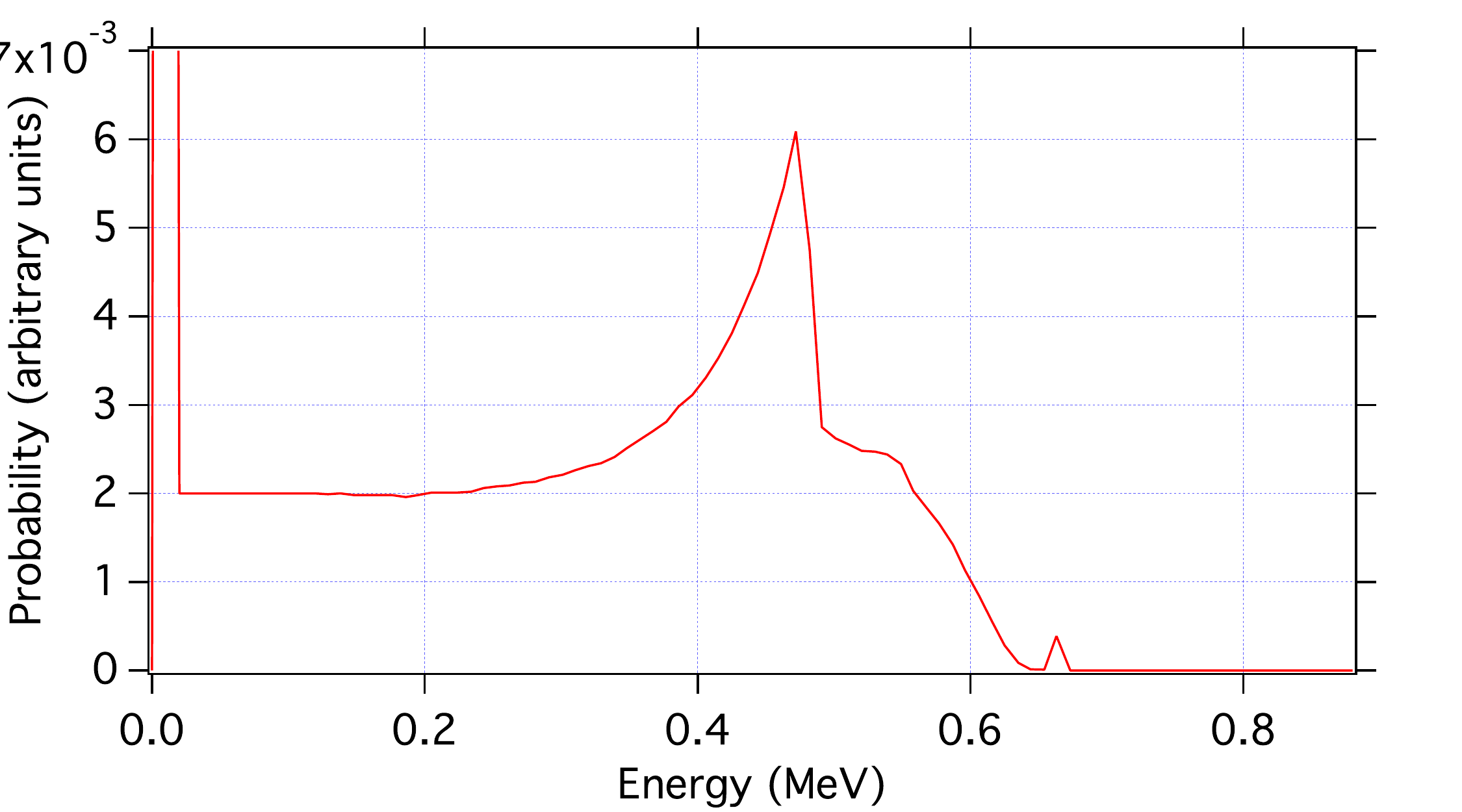} \\
 \includegraphics[width=.45\textwidth]{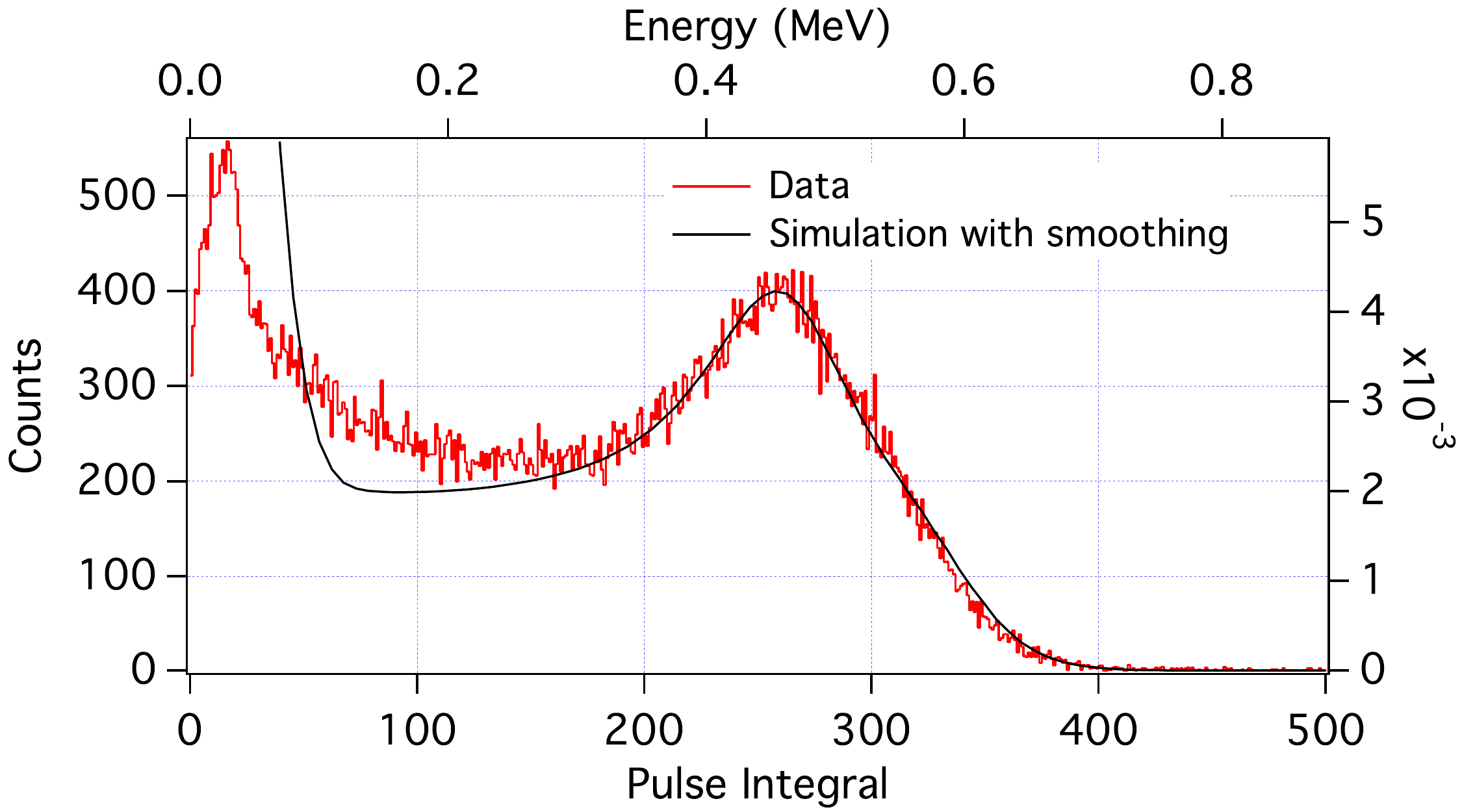}
\caption{Top: The simulated spectrum generated for a $^{137}$Cs source placed above one FaNS-1 scintillator segment. Bottom: The spectrum of deposited energy from a $^{137}$Cs gamma source.  The MCNP5 spectrum after smoothing is overlaid. }
\label{fig:gamma}
\end{center}
\end{figure}

Because plastic scintillator is a low-Z material, the dominant feature in the gamma energy spectrum is the Compton edge. To achieve an accurate calibration, the data were compared to a simulated spectrum generated by MCNP5~\cite{MCNPX2008}. The top plot of Fig.~\ref{fig:gamma} shows the MCNP5 spectrum for a $^{137}$Cs source.  To account for detector resolution, a Gaussian smoothing routine was applied, and the smoothed MCNP5 spectrum was fit the to energy region surrounding the Compton edge. This process was repeated while varying the smoothing parameters, and the fit quality was tracked by the resulting $\chi^2$. A measured $^{137}$Cs is shown in Fig.~\ref{fig:gamma} and compared with the smoothed MCNP5 spectrum.

\begin{figure}
\begin{center}
{ \includegraphics[width=.5\textwidth]{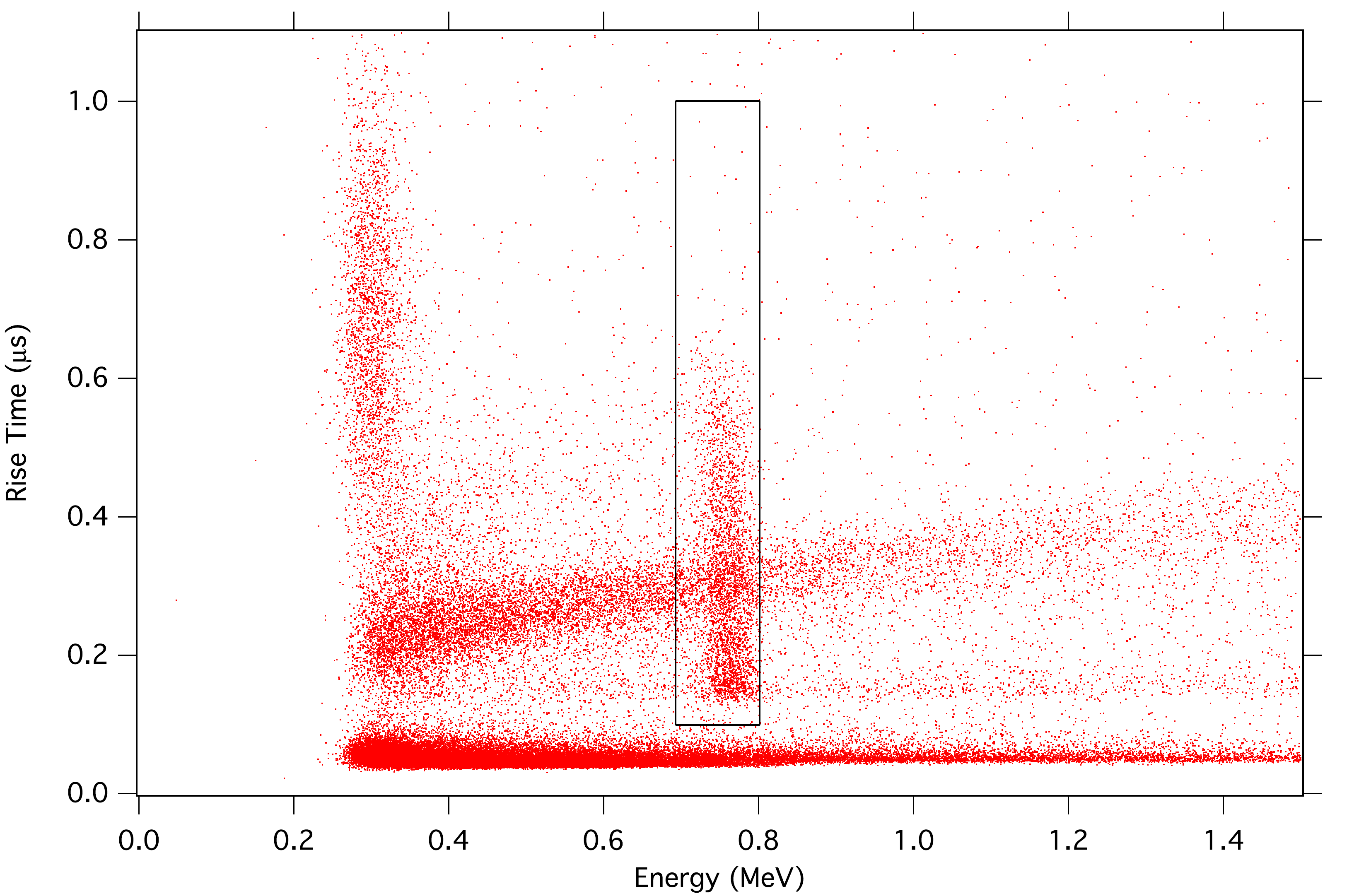}}
\caption{The 10\,\% to 50\,\% risetime versus energy from all the FaNS-1 \Het\ proportional counters for the underground data at KURF. The events in the box correspond to the full energy collection of the neutron capture reaction products.}
\label{fig:Scatter}
\end{center}
\end{figure}

The signal from a proportional counter was sent through a charge integrating preamplifier, therefore the energy of the recorded waveform is simply its amplitude. The risetime of the preamplifier signal, defined here as the time from 10\,\% of the full height to 50\,\% of the full height, is also extracted. This typically ranges from 200\,ns to 1\,$\mu$s. Figure~\ref{fig:Scatter} shows the risetime versus pulse height spectrum of a \Het\ counter for a long exposure of ambient background data taken underground (discussed in Section~\ref{sec:kurf}). The risetime is a simple measure of the geometry of the ionization trail of a neutron capture event in the \Het\ counter. A dense cluster of charge from an alpha particle, which has a large specific energy loss, creates a short risetime, while a low specific energy loss, from an electron, will leave a long track with a low total energy. Different particles cluster in different regions of risetime versus energy space, allowing for the rejection of most non-neutron capture events. A more detailed discussion of the features of the figure is found in Ref.\cite{Langford2013}

In analysis, one searches for the scintillator signal that is coincident within the $400\;\mu$s timing window of the \Het\ proportional counter signal. Events with scintillator signals that precede the \Het\ signal are considered as candidate neutron events. However, events which have a helium signal that precedes a scintillator signal are uncorrelated coincidences and constitute a flat background in the timing spectrum. This separation of true recoil-capture events and uncorrelated coincidences allows for the accurate subtraction of the background. The probability of multiple gamma or neutron events within a single timing window was small due to the low event rates encountered in these measurements. For events with multiple signals, the events were discarded with negligible effect on the response spectra. In principle, the analysis can be modified to include multiple signals to enable operation in higher rate fields.

\begin{table*}
\caption{The cut parameters for each type of data collected. The second column is the energy threshold for the combined deposition in the plastic scintillators; the third column is the energy acceptance window for the \Het\ proportional counter;  the fourth column is the range of the coincidence timing window relative to a prompt coincidence; and the final column is the range of the rise time window. N/A indicates that the cut was not applicable due to the electronics used for those measurements.}
\begin{center}
\begin{tabular}{ccccc}
\hline
Type of measurement			& Scintillator energy				& \Het\ energy 			& Coincidence			& Risetime			\\
							& threshold (MeV)				& window (MeV) 		& window ($\mu$s)		& window ($\mu$s)		\\ 
\hline
2.5\,MeV and 14\,MeV			& 0.5							& (0.1, 0.9)	 		& ($-100$, $+100$)		& N/A				\\
Surface data	 				& 1							& (0.62, 0.9)			& ($-100$, $+300$) 		& N/A				\\
Underground  data		 		& 1.1							& (0.69, 0.8) 			& ($-200$, $+200$)		& 0.1-1.0				\\
\hline
\end{tabular}
\end{center}
\label{table:params}
\end{table*}

While the rates encountered in this work were not high, one can make an estimate of the maximum background fields in which FaNS-1 can operate. For thermal neutron background, the limitation is primarily from deadtime in the data acquisition. Because of the large digitization range required for each trigger, this effect becomes noticeable at rates of around 100/s. For gamma backgrounds, a problem arises when the rate of false coincidences begins to outnumber the real coincidences. The detector has a limitation with the statistical subtraction of random coincidences when the gamma rate causes the majority of events to have multiple clusters of PMT signals. With the maximum capture window of 400\,$\mu$s, this limiting rate commences at a few 1000/s. We note that detectors based on capture-gated spectroscopy can minimize both effects by decreasing the neutron capture time, which can be accomplished in the design of the detector geometry and the material selection.

Three types of data were collected in this work: the energy response of 2.5\,MeV and 14\,MeV neutrons and the energy response of fast neutrons at the Earth's surface and underground. Analysis cuts were optimized for the data from the specific type of run, or they changed as a result of improvements to the hardware. For example, the improved electronics for \Het\ detection that were implemented for the underground data runs reduced the noise and improved the energy resolution, thus permitting a narrower energy window for the \Het\ acceptance. Table~\ref{table:params} gives a summary of some of the important cuts for the different data runs.

\section{Energy Response to Monoenergetic Neutrons}
\label{sec:DetCharac}

To understand the performance of the spectrometer, it was important to characterize it in known fast neutron fields. This was accomplished through a series of measurements performed at the Californium Neutron Irradiation Facility (CNIF) at NIST~\cite{Grundl1977}. The CNIF provided monoenergetic neutron generators at 2.5\,MeV and 14\,MeV and several \Ctft\ neutron sources whose activities were known at the level of 1\,\% to 2\,\%. The sources and monoenergetic neutrons were used to characterize the energy response.  Sealed-tube neutron generators produced the monoenergetic neutrons via the reactions deuterium-deuterium (D-D) at 2.5\,MeV and deuterium-tritium (D-T)  at 14\,MeV. 

One of the beneficial features of the CNIF is that the sources are placed in the center of a large volume room, thus minimizing the contribution of neutrons that scatter from the walls and return to the detector (often referred to as ``room return'')~\cite{Eisenhauer1989}.  
Within the main room of the CNIF,  there is an inner room whose walls, floor, and ceiling are constructed of thin-wall aluminum that is 5\,cm thick and filled with anhydrous borax. This provides good thermal neutron shielding and shields the detector from neutrons that thermalize outside the inner structure. During irradiations, the detector was placed within the inner room near a neutron source, and the resulting fast neutron field was a combination of an isotropic distribution from the source and neutron return from the environment, including albedo from boundaries of the irradiation room.

\begin{figure}[t]
\begin{center}
\includegraphics[width=.45\textwidth]{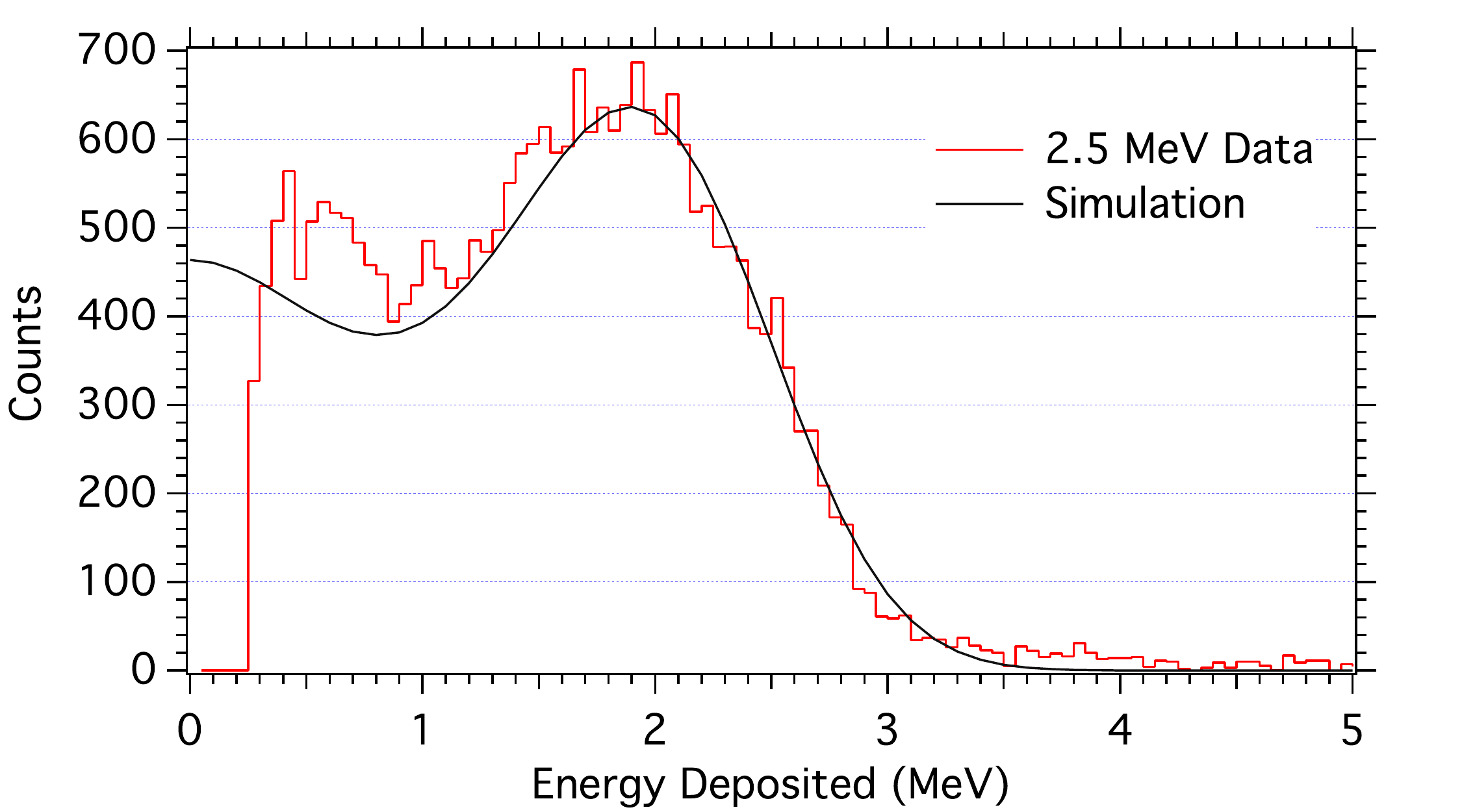} \\
\includegraphics[width=.45\textwidth]{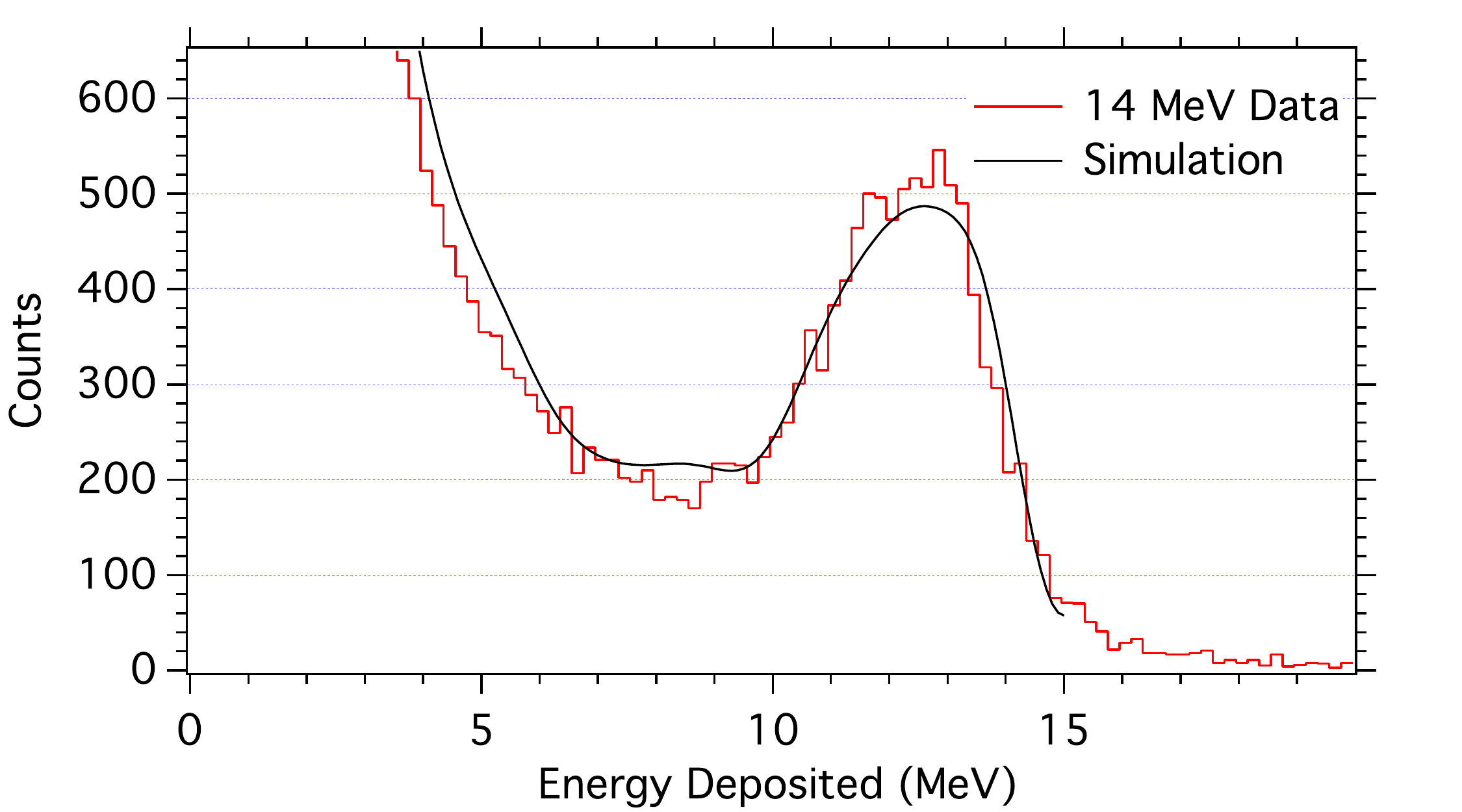}
\caption{Pulse height spectra from two monoenergetic neutron generators. Top: Response from a 2.5\,MeV neutron generator. Bottom: Response from a 14\,MeV neutron generator. The MCNPX simulation is overlaid on each plot.}
\label{fig:monospec} 
\end{center}
\end{figure}

To obtain the energy response, the spectrometer was irradiated with both 2.5\,MeV and 14\,MeV neutrons.  The data were analyzed as discussed in Section~\ref{sec:Analysis} to yield pulse height spectra. To obtain the true energy conversion, the detector was calibrated for energy using the Compton edges of the gamma rays from $^{60}$Co and $^{137}$Cs. This gives a reliable calibration in electron equivalent energy. The pulse height data from each of the detector sections was then converted to energy using the light response of Eqn.~\ref{eqn:Birks} and added together to give the total energy of an event. Figure~\ref{fig:monospec} shows the resulting energy response for the 2.5\,MeV and 14\,MeV neutrons. The agreement between data and simulation indicates that the light response function of Eqn.~\ref{eqn:Birks} reproduces the observed neutron spectra well. This is an important result because the same function will be used for determining the energy response of the data for the surface and underground measurements. In addition, the spectra exhibit well-resolved peaks without using any unfolding procedure. This highlights a benefit of capture-gated spectroscopy; with the FaNS-1 detector, one is able to discern directly the energy response of monoenergetic sources. The low-energy portions of the plot are populated by room-return neutrons that have lost significant energy in the surrounding environment.

The FaNS-1 spectrometer was modeled with both MCNP5 and MCNPX~\cite{MCNPX2008,MCNPX2011}. MCNP5 was adequate for gamma-ray interactions, but for higher energy neutrons, the simulations were performed in MCNPX, which extends to higher energies than MCNP5 and allows tracking of radiation other than neutrons, gamma rays, and electrons. MCNPX allows some variation in the physics used for the calculations. For this work, the MCNPX tabular physics were used below 20\,MeV, and the default model physics were used above.

Calculations of the neutron response of the detector used the PTRAC option of MCNPX, which allows the proton, deuteron, triton, and alpha particles to be tracked separately. The proton-recoil light response discussed in Section~\ref{subsec:energy} was used to convert deposited energy into light on an event-by-event basis. The light response for deuterons, tritons, \Het, and \Hef\ was derived from the proton response. The light output $L$ depends mainly on the magnitude of the differential energy loss, which depends on the velocity and charge of the particle. Thus, the following scaling was used for particles of equal charge $Z$ and equal energy per nucleon, $E/A$, but different mass numbers $A1$ and $A2$

\begin{equation}
\label{eqn:breuer}
L\Big(A_2, {E_2\over{A_2}}\Big) = L\Big(A_1, {E_1\over{A_1}} \Big) \times {A_2\over{A_1}},\ \text{for}\ Z_1 = Z_2.  \\
\end{equation}

Numeric filters were developed to analyze the PTRAC files and report the time between the first collision and capture, whether there was a \Het\ capture, and total light produced in individual scintillators. These filters allowed for the matching of experimental parameters in the simulation.

When environmental scattering (e.g., from floors or walls) was thought to be significant, the FaNS-1 model was inserted into various geometries. The effects of gamma rays were also investigated, and the study included background gamma rays, gamma rays from sources used in calibration, and gamma rays resulting from neutron interactions. The latter effects were found to be quite small.

\section{Surface Measurement}
\label{sec:surface}

The surface flux and energy spectrum of fast neutrons have been studied extensively as a function of altitude and latitude~\cite{Goldhagen2002,Gordon2004}. These measurements typically use Bonner spheres, which rely on unfolding the energy spectra from multiple detectors and assumptions of the underlying spectrum. The method of capture-gated spectroscopy provides a path for improvement by directly recording the energy deposition from incident neutrons. Nevertheless, it still must address the complications arising from the various energy loss mechanisms as well as understanding the light response function of the scintillator over a large energy range.

A simulation of FaNS-1 in an isotropic neutron field was performed to determine the detector sensitivity. The input energy distribution came from the work of Ref.~\cite{Gordon2004}, which is currently the JEDEC standard cosmogenic neutron spectrum~\cite{JESD89a}. The JEDEC spectrum is based on measurements performed above a hard ground and therefore includes backscattered neutrons. This causes a significant enhancement of the spectrum for neutrons below 100\,MeV. Therefore, the simulation used an isotropic field of neutrons, without ground, to avoid double-counting backscattered neutrons.  Simulated events must pass analysis cuts on the deposited energy in each detector segment and time between neutron scatter and capture. In this way the simulation was matched to experimental parameters. Two correction factors were applied: one of $0.770 \pm 0.015$ that accounts for the efficiency of the energy cut on \Het\ captures (discussed in Section~\ref{sec:Analysis}), and second of $0.85\pm0.15$\ that accounts for the intrinsic efficiency of our \Het\ proportional counters. With this result, one may define an average sensitivity $\overline{S}^{surf} = 8.1\pm 1.4$~n/(n/cm$^2$) for cosmogenic neutrons of 1\,MeV and above. This sensitivity can be interpreted as an efficiency weighted by the input energy spectrum. 

To measure the surface neutron flux and energy response, the detector was placed in a steel cargo container on the grounds at NIST. The position of the detector within the cargo container was approximately six meters away from a six meter tall building. There was no additional shielding near the detector. The location was 39.130$^{\circ}$\,N, 77.218$^{\circ}$\,W at an elevation of 120 meters above sea level, and the detector was 1.5\,m above ground. Data were collected at the end of May 2010. During two days of operation, the detector was live for 45 hours.

\begin{figure}[h]
\begin{center}
\includegraphics[width=.5\textwidth]{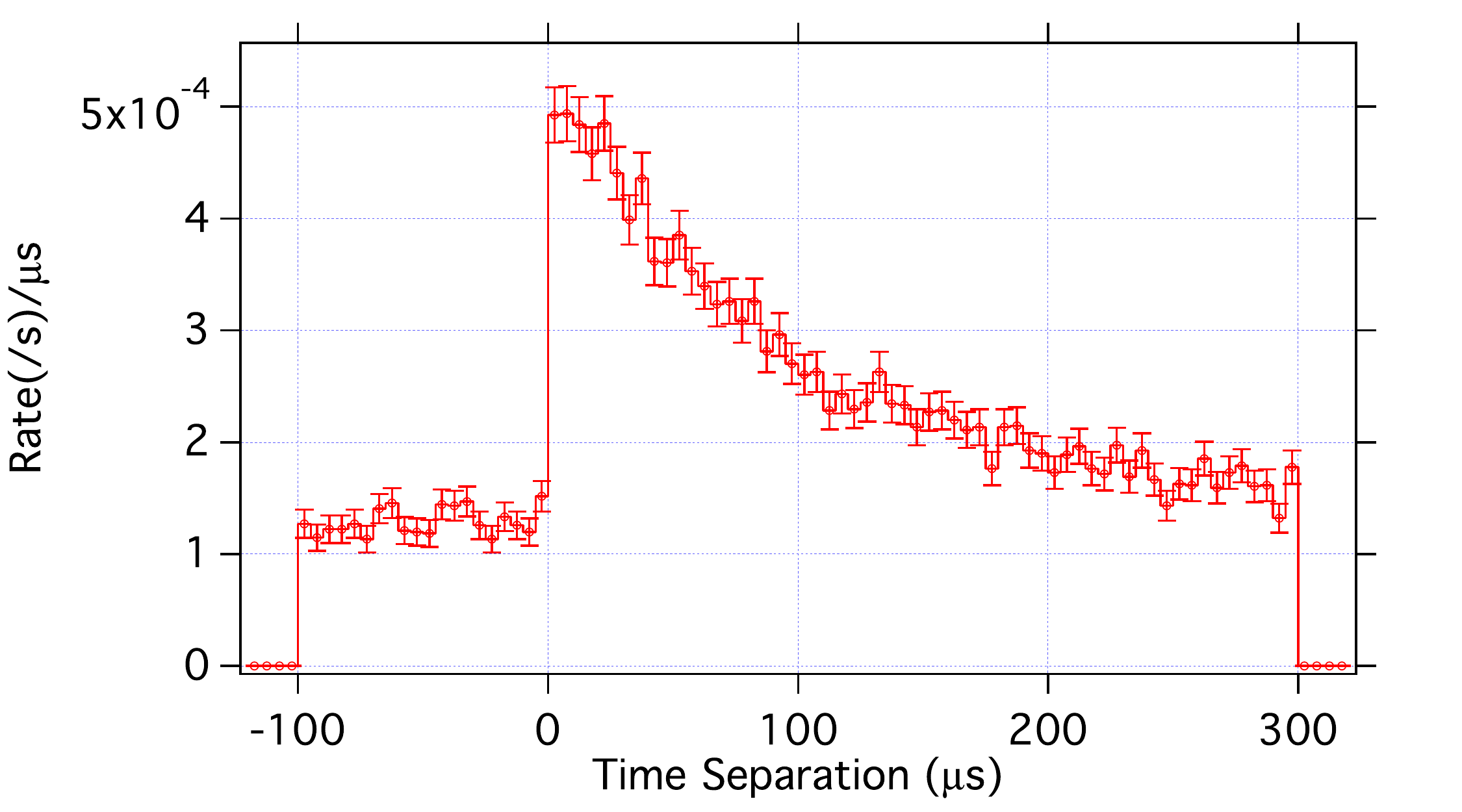}
\includegraphics[width=.5\textwidth]{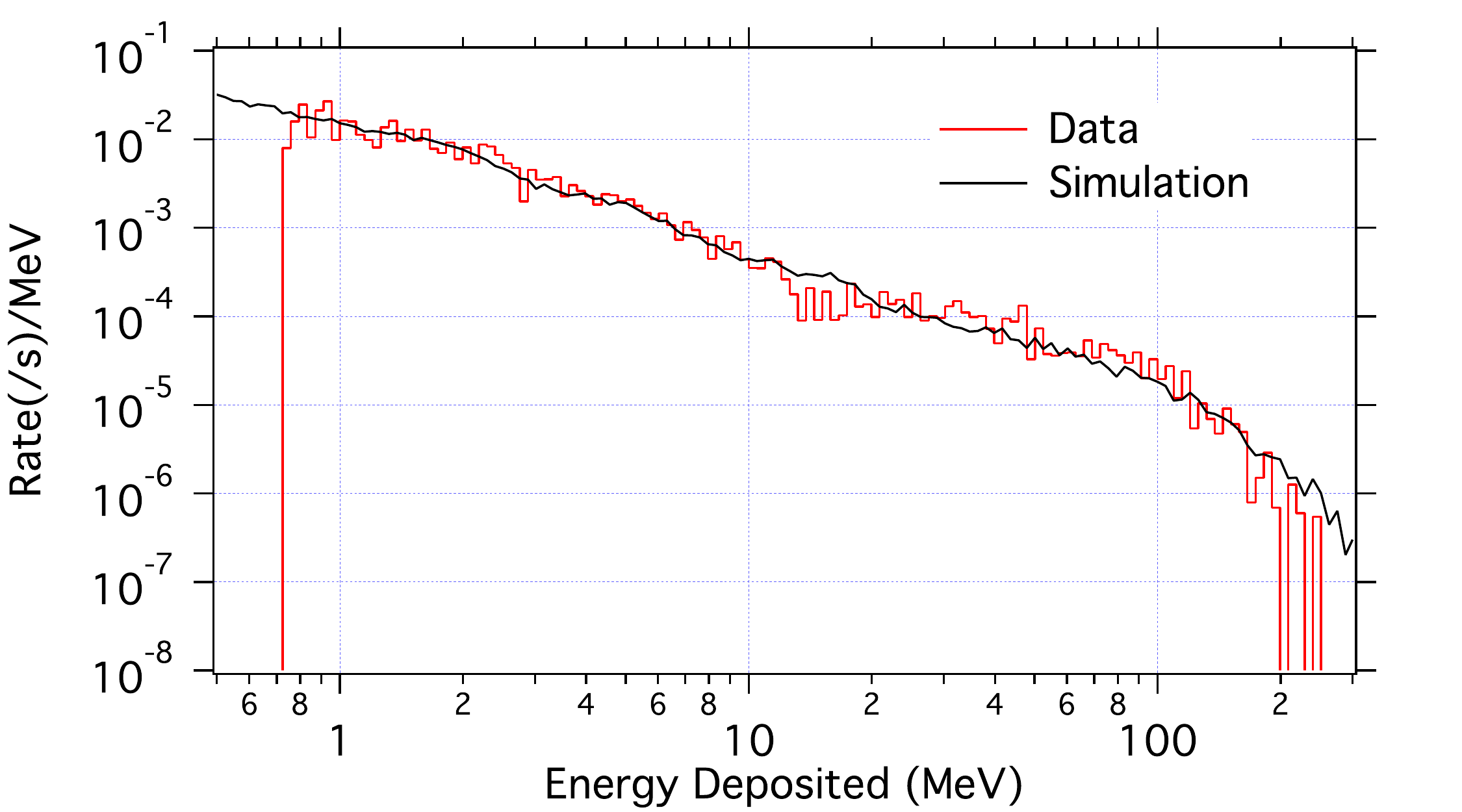}
\caption{Top: The timing spectrum from the data run on the surface at NIST Gaithersburg. The error bars are statistical only. Bottom: The detected neutron energy response at NIST Gaithersburg after background subtraction. Overlaid is an MCNPX simulation of the detector's response to the reported spectrum from Ref.~\cite{Gordon2004}. The absolute count rate normalization between the data and simulation has been allowed to float to account for differences in the environmental conditions between our data and the data of Ref.~\cite{Gordon2004}.}
\label{fig:surface} 
\end{center}
\end{figure}

Approximately $1.31\times10^5$ events were recorded during an exposure time of $\tau = 1.62\times10^5$\,s, giving a total trigger rate of 0.8\,s$^{-1}$. After applying coincidence requirements and cuts  from Table~\ref{table:params} on the neutron capture energy in the helium detectors, $1.10\times10^4$ events remained, giving a post-cut rate of 0.068\,s$^{-1}$. The number of neutrons remaining after subtracting random coincidences is $N=5133\pm151$, yielding a final neutron count rate of 0.032\,s$^{-1}$. The timing distribution is shown in Fig.~\ref{fig:surface}. For this data set, the acquisition time window was extended to improve the coincidence efficiency. Because the detector operated at a low trigger rate, this did not increase deadtime.

Using the same background subtraction and pulse height method as before, we obtained the neutron energy deposition seen in Fig.~\ref{fig:surface}. The detected neutron flux above 1\,MeV at the surface $\Phi_n^{surf}$ using the simulated sensitivity $\overline{S}^{surf}$ yields

\begin{align}
\nonumber
\Phi_n^{surf} 	&= \frac{N}{\tau \times \overline{S}^{surf}}\\
			&= (3.9\pm 1.1)\times10^{-3}\,\textrm{\,cm$^{-2}$ s$^{-1}$}.
\end{align}

\noindent Table~\ref{tab:uncertainty} gives a summary of the statistical and systematic uncertainties that comprise the total uncertainty. The largest contribution comes from a 20\% uncertainty in the knowledge of the scintillator energy threshold.

\begin{table*}
\caption{Summary of correction factors and relative uncertainties for the measured fast neutron fluxes at the surface and underground at KURF. The uncertainties are associated with systematic effects with the exception of those for the counting and simulation statistics. An entry of zero means that no explicit correction was made.} 
\begin{center}
\begin{tabular}{lrrrr}
\hline
						& Surface		&					&KURF			\\
\hline
Source					&Correction	&Relative Uncertainty	&Correction	&Relative Uncertainty	\\
\hline
Energy threshold			& 0			& 20\%				& 0			& 20\%	\\
Counting statistics			& 0			& 3\% 				& 0			& 21\%	\\
\Het\ counter efficiency 		& 0.85		& 18\% 				& 0.85		& 18\%	\\
\Het\ cut efficiency			&0.77 		& 2\% 				& 0.77		& 2\%	\\ 
Simulation statistics			& 0 			& 5\%				& 0			& 5\%	\\
Exposure	time				& 0 			& 1\%				& 0			& 1\%	\\
\hline
Total						&			&28\%				&			&35\%	\\			
\hline
\end{tabular}
\end{center}
\label{tab:uncertainty}
\end{table*}

A comparison between the simulated and measured neutron response is shown in Figure~\ref{fig:surface}. For this comparison, we let the absolute normalization of the spectrum~\cite{Gordon2004} float because the detector operated in different environmental conditions (solar cycle, barometric pressure, humidity, latitude, etc.) that can effect the absolute neutron flux. The measurement at the surface verifies that the response of the detector was modeled correctly in MCNPX and that the spectrum of Ref.~\cite{Gordon2004} is an appropriate choice for the simulation. This gives confidence that the simulation can be used to infer the neutron flux in a low neutron flux environment.

\section{Underground Measurement}
\label{sec:kurf}

After the completion of the calibration runs and the acquisition of the surface data, the spectrometer was moved to the Kimballton Underground Research Facility (KURF), located at Lhoist North America's Kimballton mine in Ripplemead, VA. The purpose was to measure the ambient fast neutron response at KURF~\cite{Best2012} and also to characterize the detector in a low-background environment. The facility is located in an active limestone mine at a depth of 1450 meters of water equivalent (mwe) and provides a good low-radioactivity counting environment~\cite{Finnerty2010}. The FaNS-1 detector resided directly on the concrete floor of the KURF laboratory for all of the measurements discussed in this work.

\begin{figure}[t]
\begin{center}
\includegraphics[width=.5\textwidth]{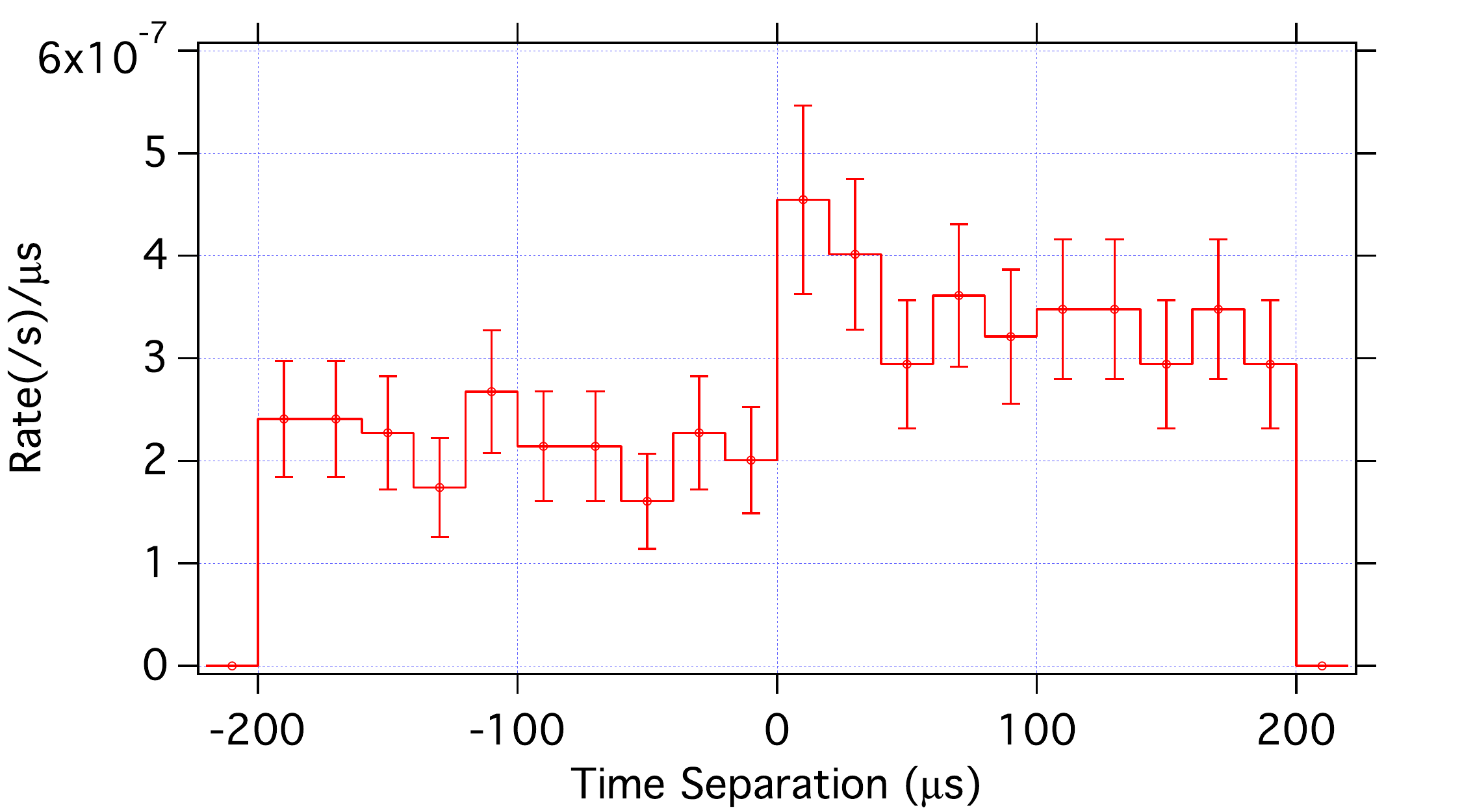}
\includegraphics[width=.5\textwidth]{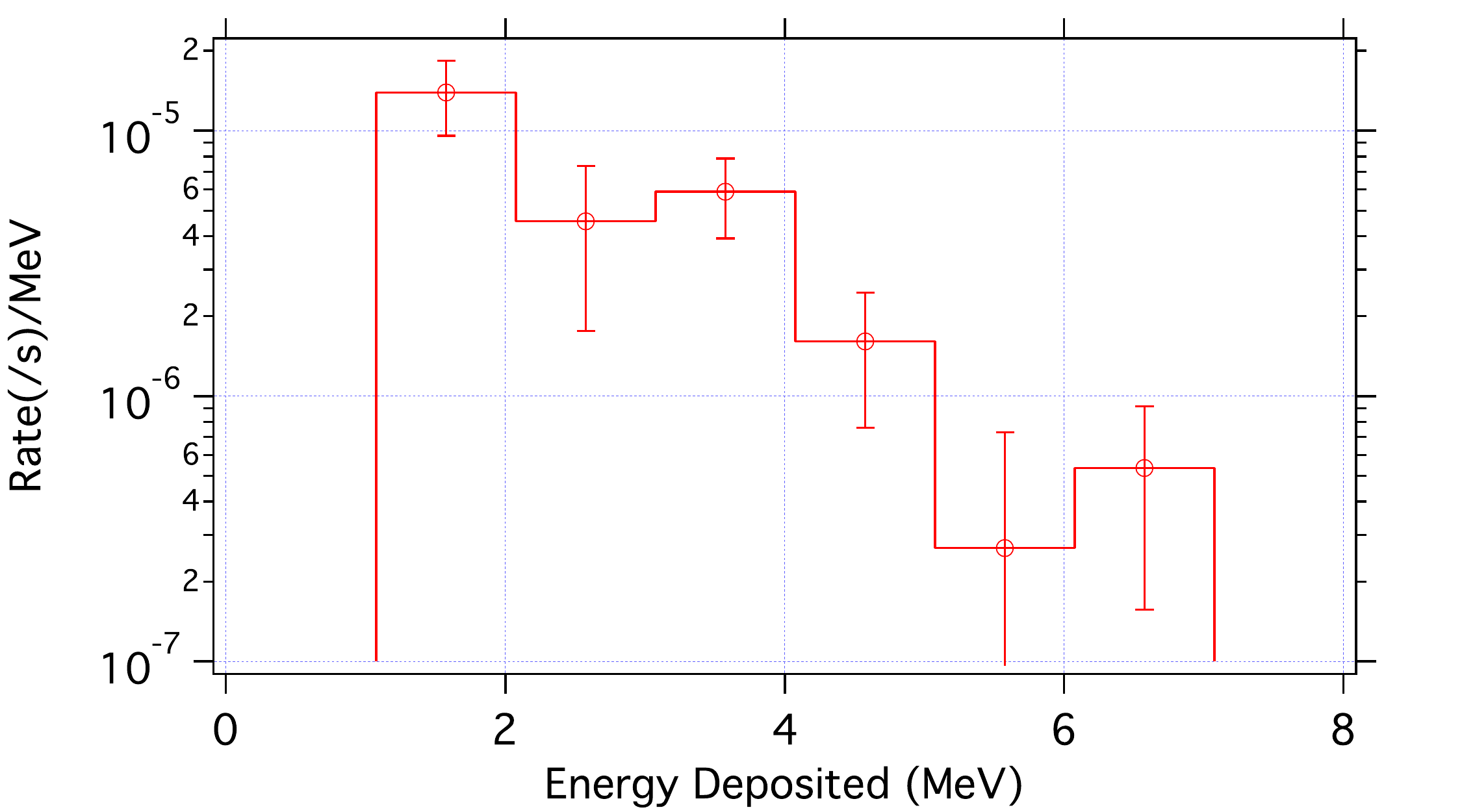}
\caption{Top: The timing spectrum from the 2 month exposure at KURF. Bottom:  The detected neutron energy response at KURF after background subtraction. The error bars in both plots are statistical only.}
\label{fig:KURF}
\end{center}
\end{figure}

The detector commenced operation at KURF in the summer of 2010. In approximately 2 months of operation with an exposure time of $\tau= 3.737\times10^6$\,s, 434 events with energies greater than 1.1\,MeV were detected. The data were analyzed in the same manner as the characterization data and the surface data, and the cut parameters are given in the Table~\ref{table:params}. The timing spectrum for the underground measurement is shown in Fig.~\ref{fig:KURF}. Of the total number of events in the timing spectrum, 272 have positive timing differences, and 162 have negative timing differences, and thus 110 events are attributed to neutron capture. 

There is a small systematic effect seen in the timing spectrum that is caused by prompt alpha-gamma coincidences. It occurs when the decay of a radioactive contaminant from the aluminum body of the \Het\ counter produces an alpha and gamma in coincidence~\cite{Langford2013}. These appear in the spectrum as valid events: a \Het\ capture in coincidence with a scintillator signal. To correct for this, the timing spectrum was rebinned in microsecond increments that revealed an anomalously high 2-$\mu$s wide peak near $t=0$ attributable to these coincidences. That peak had 10 events and was subtracted from the 110 neutron capture events. By making this correction, a small number of real neutron captures may also have been rejected. We estimate this fraction to be approximately 2\,\% and do not make any correction for it. Note that this effect is also in the surface data, but its contribution is negligible because of the higher neutron rate.

After the uncorrelated background subtraction and the removal of prompt events, a total of $N=100 \pm 21$  events were observed. The energy response and timing spectra of these data are shown in Fig.~\ref{fig:KURF}. To quantify the detector response to the underground neutron flux, a simulation was performed using a uniform distribution of ($\alpha$,n) neutrons incident on FaNS-1. The dominant source of 1-10\,MeV neutrons underground is ($\alpha$,n) interactions in the surrounding material, but these interactions are complicated and depend on the isotopic abundance and distribution in the material. Without performing detailed analysis of the materials composition in the surrounding environment, a precise spectrum cannot be predicted. Therefore, to approximate the (alpha, n) spectrum, we used a standard AmBe spectrum~\cite{IAEA2001} for the simulation. As check on the sensitivity to choice of the input spectrum, we also simulated a flat spectrum from 1-8\,MeV (the range of neutrons detected at KURF) and found only a small effect on the value of $\overline{S}$.

After applying the same method as in Section~\ref{sec:surface}, we obtain an average sensitivity to these neutrons of $\overline{S}^{KURF} = 15.1\pm 2.9$~n/(n/cm)$^2$ and determine the observed flux above 1.1\,MeV at KURF to be

\begin{align}
\nonumber
\Phi_n^{KURF}	&= \frac{N}{\tau \times \overline{S}^{KURF}}\\
		&= (1.8\pm 0.6)\times10^{-6}\,\textrm{\,cm$^{-2}$ s$^{-1}.$}
\label{eqn:KURF}
\end{align}

\noindent Table~\ref{tab:uncertainty} gives a summary of the statistical and systematic uncertainties that comprise the total uncertainty. This flux is comparable to measurements and simulations of the ($\alpha$, n) neutron flux in other underground labs~\cite{Chazal1998,Carson2004}. No attempt was made to measure the intrinsic background of the detector, but based on its material composition and an estimate of the intrinsic radioactivity, the contribution is expected to be very small.

One may obtain an estimate of the total muon-induced neutron flux using the method of Ref.~\cite{Mei2006}. Their formula for the neutron flux as a function of depth is

\begin{equation}
\phi_n = P_0 (P_1/h_0) e^{-h_0/P_1}, 
\label{eqn:Mei}
\end{equation}

\noindent where $h_0$ is the vertical depth in kilometer of water equivalent (km.w.e.).  The fit parameters are $P_0 = (4 \pm 1.1) \times 10^{-7}$\,cm$^{-2}$s$^{-1}$ and $P_1 = (0.86 \pm 0.05)$\,km.w.e.  At the depth of KURF, one obtains an expected total muon-induced neutron flux of $3.4\times10^{-8}$\,cm$^{-2}$s$^{-1}$. The authors predict that 22\% of these neutrons are above 10~MeV, which yields a flux above 10~MeV of $7.5\times10^{-9}$\,cm$^{-2}$s$^{-1}$.

No events were observed with energies greater than 10\,MeV during the period of operation of FaNS-1 at KURF. Other work has shown that muon-induced neutrons above 10\,MeV should have a similar shape to cosmogenic neutrons at the surface although with a substantially reduced flux. One can estimate an average sensitivity to these neutrons from a simulation of the cosmogenic neutron spectrum. For FaNS-1 we determine an average sensitivity  for muon-induced neutrons with energies above 10\,MeV to be $\overline{S}^{KURF} = 3.5\pm 0.7$~n/(n/cm$^2$). Using that result and a value for $N^{KURF} =3$ at the Poisson 95\% confidence level,  Eqn.~\ref{eqn:KURF} yields an upper limit of the neutron flux above 10\,MeV of

\begin{align}
\Phi_n^{KURF}(>10\,{\rm MeV}) \leq 2.3\times10^{-7}~\textrm{\,cm$^{-2}$ s$^{-1}$}.
\end{align}

\noindent This upper limit is considerably higher than the expected rate calculated from Eqn.~\ref{eqn:Mei}. If one extrapolates from our simulated sensitivity, one would expect FaNS-1 to observe only about one muon-induced neutron per year of operation at KURF. Such a measurement is not practical using FaNS-1 at KURF, and therefore one must consider moving to a shallower depth or constructing a significantly larger detector to study the cosmically-induced part of the energy spectrum. 

\section{Summary and Outlook}
\label{sec:summary}

We constructed a fast neutron spectrometer consisting of plastic scintillator and \Het\ proportional counters. The detection principle uses segmentation to improve energy resolution and capture-gated spectroscopy to reduce backgrounds. The spectrometer is capable of observing very low neutron fluxes in the presence of ambient gamma background, and this technique does not require scintillator pulse-shape discrimination. Using monoenergetic neutron generators and calibrated neutron sources, we characterized the neutron response of the detector. The results compare very favorably with simulations performed with MCNP. Additional work is required to extract the incident kinetic energy of the incident neutron from the measured detector response.

A measurement of the surface neutron flux at NIST in Gaithersburg, MD was made of $(3.9\pm 1.1)\times10^{-3}$\,cm$^{-2}$ s$^{-1}$ for energy above 1\,MeV. The data agree reasonably well with a simulated spectrum based on another measurement. The spectrometer was subsequently installed at KURF and measured the ambient fast neutron flux above 1.1\,MeV. The measurement yielded a result of $(1.8\pm 0.6) \times 10^{-6} $\,cm$^{-2}$ s$^{-1}$. 

It is important to understand the flux and energy spectrum of muon-induced fast neutrons. The depth of KURF and the efficiency of FaNS-1 detector make it impractical to improve on these measurements. Improvement would come from performing the measurements at a shallower depth, and increasing both the efficiency and energy range of the spectrometer. Toward that end, this collaboration has developed a second detector, FaNS-2, that has an energy range of 1\,MeV to 1\,GeV. The detector uses the same principles of capture-gated spectroscopy and segmentation as FaNS-1 to achieve good background rejection and energy resolution. It is a larger volume detector composed of 16 bars of plastic scintillator with 21 \Het\ proportional counters interspersed among them. The geometry was optimized using MCNPX simulations. Its sensitivity to cosmogenic neutrons is approximately 90 n/(n/cm$^2$)~\cite{LangfordThesis} versus 8.1 n/(n/cm$^2$) for FaNS-1.

This increased sensitivity along with deploying the detector at shallower depth should permit the acquisition of an energy spectrum with significantly higher statistics. Two of the main challenges will be understanding the sources of background at high energy where the event rate is low and determining the light response function of the scintillator over the large energy range. 

\section{Acknowledgments}

We thank Vladimir Gavrin and Johnrid Abdurashitov of the Institute for Nuclear Research - Russian Academy of Sciences for useful discussions. We acknowledge the NIST Center for Neutron Research for the loan of the \Het\ proportional counters used in this work. We also acknowledge KURF and Lhoist North America, especially Mark Luxbacher, for providing us access to the underground site and logistical support. The research has been partially supported by NSF grant 0809696. T. Langford acknowledges support under the National Institute for Standards and Technology American Recovery and Reinvestment Act Measurement Science and Engineering Fellowship Program Award 70NANB10H026 through the University of Maryland.

\newpage
\bibliographystyle{elsarticle-num}
\bibliography{FNSurface.bib}
\end{document}